**Saggu, A., & Ante, L. (2023). The Influence of ChatGPT on Artificial Intelligence Related Crypto Assets: Evidence from a Synthetic Control Analysis. Finance Research Letters, 103993.**

# The Influence of ChatGPT on Artificial Intelligence Related Crypto Assets: Evidence from a Synthetic Control Analysis

Aman Saggu[ab]*

[a]*Business Administration Division, Mahidol University International College*

[b]*Blockchain Research Lab*

Lennart Ante[b]

[b]*Blockchain Research Lab*

**Abstract:**

The introduction of OpenAI's large language model, ChatGPT, catalyzed investor attention towards artificial intelligence (AI) technologies, including AI-related crypto assets not directly related to ChatGPT. Utilizing the synthetic difference-in-difference methodology, we identify significant "ChatGPT effects" with returns of AI-related crypto assets experiencing average returns ranging between 10.7% and 15.6% (35.5% to 41.3%) in the one-month (two-month) period after the ChatGPT launch. Furthermore, Google search volumes, a proxy for attention to AI, emerged as critical pricing indicators for AI-related crypto post-launch. We conclude that investors perceived AI-assets as possessing heightened potential or value after the launch, resulting in higher market valuations.

*Corresponding author: Aman Saggu, Business Administration Division, Mahidol University International College, 999 Phutthamonthon Sai 4 Rd, Salaya, Phutthamonthon District, Nakhon Pathom 73170, Thailand, Telephone: 0066 27005000, Fax: 0066 24415091, Email address: aman.sag@mahidol.edu; asaggu26@gmail.com

**Acknowledgements**

We are grateful for helpful comments and suggestions to the editors Laura Ballester; Jonathan A. Batten; and Samuel Vigne; anonymous referees; and the Blockchain Research Lab.

**Funding**

This research did not receive any specific grant from funding agencies in the public, commercial, or not-for-profit sectors.

**Conflict of Interest Statement**

The authors whose names are listed immediately below certify that they have NO affiliations with or involvement in any organisation or entity with any financial interest (such as honoraria; educational grants; participation in speakers' bureaus; membership, employment, consultancies, stock ownership, or other equity interest; and expert testimony or patent-licensing arrangements), or non-financial interest (such as personal or professional relationships, affiliations, knowledge or beliefs) in the subject matter or materials discussed in this manuscript.

Aman Saggu

Lennart Ante

# 1 Introduction

The launch of OpenAI's ChatGPT on November 30, 2022, marks a significant milestone in the advancement of Artificial Intelligence (AI).[1] ChatGPT is a state-of-the-art transformerbased large language model with sophisticated natural language processing capabilities, enabling real-world applications across diverse domains. Trained on a vast textual corpus, the ChatGPT model can generate human-like responses to user inputs known as prompts (OpenAI, 2022). ChatGPT exemplified its potential through superior performance in a Google interview for a highly sought-after role (Bhaimiya, 2023) and successful completion of examinations for law and business schools (Murphy Kelly, 2023). Its utility has been demonstrated by its impressive popularity, exceeding 100 million active users in January 2023, establishing a record for the fastest-growing user base in history (Savitz, 2023).[2]

The ground-breaking ChatGPT technology has stimulated commercial AI development (Tyrrell, 2022) and catalyzed digitalization initiatives (Brown, 2022). Media coverage predicts integration of ChatGPT into Microsoft's Bing search engine could position it as a formidable online search market competitor, compelling other tech giants like Google and Baidu to prioritize AI development (Elias, 2023). These companies' reactions to ChatGPT signaled an elevated perceived value of AI technology among investors. For example, c3.ai, a software company specializing in AI, saw its share price surge 28% following the announcement of ChatGPT integration into its suite of AI tools (Fox, 2023).[3] The influence of ChatGPT transcends directly related companies to the wider AI sector, as evidenced by the rising share prices of BigBear.ai and SoundHound AI following the ChatGPT launch, despite their lack of association to ChatGPT (Singh and Biwas, 2023), signaling that investors perceived AI-related companies as possessing heightened potential or value. Moreover, empirical research has further shown that AI-focused exchange-traded funds exhibited a name-premium of around 0.4% even before the ChatGPT launch, indicating that investors recognized the growing importance of AI as a valuable technology (Wu and Chen, 2022).

---

[1] Large language models are deep neural network algorithms that process and generate natural language text.

[2] Despite these accomplishments, OpenAI CEO Sam Altmann cautions against overreliance on ChatGPT for critical applications as the technology has not reached full maturity (Altman, 2022).

[3] Similarly, Buzzfeed, a digital media company unrelated to AI, experienced a 120% surge in its share price after announcing OpenAI technology adoption for content creation (Diaz and Smith, 2023).

Previous academic studies have examined ChatGPT's significance in different fields, including research (Dowling and Lucey, 2023; Thorp, 2023; Zaremba and Demir, 2023), software engineering (Sobania et al., 2023; Stokel-Walker, 2023), and education (Susnjak, 2022). However, despite growing interest, there is a noticeable lack of empirical studies exploring the potential impact of ChatGPT and its subsequent hype on AI-related financial assets. A salient study revealed 90% positive abnormal returns for a select group of AI-themed cryptocurrencies (Ante and Demir, 2023).

In light of this context, this paper aims to evaluate the impact of the ChatGPT launch on the narrative trading of AI-related crypto assets (hereafter referred to as AI-assets). This focus is crucial because, while some companies listed on the stock market may integrate AI into their offerings, only a limited number specialize in AI product and service provision. In contrast, a distinct cohort of crypto assets prioritize AI as a fundamental aspect of operations.[4] Consequently, this study investigates how the introduction of ChatGPT influenced price discovery and market perceptions of AI-crypto-assets. In pursuance of this objective, we empirically investigate if: (1) returns of AI-assets and non-AI-assets were equivalent before the ChatGPT launch, (2) a disparity in investor perception between AI and non-AI-assets post-launch, and (3) this divergence can be explained by changes in Google search volumes (financial news wires) as a proxy for retail (institutional) investor attention towards AI.

---

[4] AI-centric cryptocurrencies such as SingularityNET (AGIX) and Fetch.AI (FET) appreciated 20% and 50% respectively following the ChatGPT launch, compared to just 1% and -3% changes for widely recognized benchmarks Bitcoin (BTC) and Ethereum (ETH) respectively. Calculated using CoinGecko price data between November 28, 2022, and December 12, 2022.

Firstly, we utilize the synthetic difference-in-difference (SDID) approach of Arkhangelsky et al. (2021) and Clarke et al. (2023) to examine the relative evolutionary performance of a large set of AI-assets relative to controls.[5] Results demonstrate that AI-assets exhibited positive average treatment effects (ATTs) of 10.7% to 15.6% (35.5% to 41.3%) in the one-month (two-month) period following the ChatGPT launch. This positive effect is more meaningful in context, as the broader cryptocurrency market was characterized by a bear market and extreme investor uncertainty during that time, emphasizing the significant impact of ChatGPT on the narrative trading of AI-assets.[6] Secondly, we reveal a spillover effect from attention generated by ChatGPT to AI-assets that have no direct relation to ChatGPT, contributing to ongoing research exploring the role of sentiment and hype in cryptocurrency markets (Nepp and Karpeko, 2022; Shahzad et al., 2022; Subramaniam and Chakraborty, 2020) as well as the literature on quality signals in price promotion and crypto market efficiency (Ante, 2023; Brunnermeier, 2005). Thirdly, a panel fixed-effects model furnishes empirical support for Google search volume data via Google Trends as a proxy for retail investor attention, serving as reliable price indicators for AI-assets only after the launch, highlighting the importance of incorporating such data into market analyses for effective decision-making. Lastly, we present empirical evidence that financial news wires, as a proxy for institutional investor attention, exerted limited influence on AI-crypto assets following the launch.

---

[5] Controls include non-AI-assets and benchmark crypto indices, before and after the ChatGPT launch

[6] According to the alternative.me Crypto Fear and Greed Index, the cryptocurrency market experienced a period of "fear" and "extreme fear" from November 30, 2022 to January 14, 2022. The market then fluctuated between "fear", "neutral" and "greed" on a daily basis thereafter. Additionally, the Sentix Bitcoin Sentiment Indicator revealed a bearish or neutral sentiment throughout most of this period (Sentix, 2023).

From a theoretical standpoint, there are several potential channels contributing to the positive price effect for AI-assets post-ChatGPT launch. Firstly, market efficiency theories assert that asset valuations are linked to perceived quality (Fama, 1970; Lo, 2004), and the comprehensive media coverage of the ChatGPT launch may serve as a quality proxy, leading retail investors to value AI-assets more highly in accordance with signaling theory (Spence, 1973). This aligns with the notion that retail investors, prone to cognitive biases and heuristics, tend to overreact to news or events (Ante, 2023). Secondly, institutional investors, acknowledging ChatGPT's potential and influenced by similar market psychology, may have directed capital toward AI-related projects, however institutional interest in cryptocurrencies waned. Thirdly, ChatGPT may have fostered information diffusion effects within the cryptocurrency market by equipping retail investors with the capacity to distill complex and technical concepts, thus facilitating more informed investment decisions (Binance, 2023). Finally, we acknowledge network effects arising from ChatGPT's launch may have spurred a self-sustaining growth cycle for AI-assets, potentially attracting more users and boosting demand and valuations.

## 2  Data

Our dataset spans October 1, 2022, to January 31, 2023. Asset returns, $r_t$, are defined as the first difference of the natural log of the asset price, $p_t$, relative to the previous day: $r_t = \ln(p_t/p_{t-1})$. Panel A in Table 1 presents descriptive statistics for two cohorts of crypto asset returns: a curated cohort of 16 crypto assets classified as 'AI'-related by coingecko.com (GAI) and a broader cohort of 86 crypto asset returns categorized as 'AI and Big Data'-related by coinmarketcap.com (CAI). These cohorts include a diverse range of assets, such as SingularityNET, an AI service marketplace; Numeraire, a blockchain-based hedge fund that uses AI predictions for investments; and Fetch AI, a platform focused on AI data connectivity.

---INSERT TABLE 1---

Corresponding to Panel A, Panel B presents descriptive statistics for the returns of control groups GCKO for GAI (CMC for CAI) comprising the 15 (85) largest crypto assets ranked by market capitalization on coingecko.com (coinmarketcap.com), respectively, excluding stablecoins and assets belonging to Panel A.[7] To account for potential subjectivity in asset selection, we also employ S&P Cryptocurrency BDM Ex-LargeCap Index (SPCBXL) and S&P Cryptocurrency BDM Ex-MegaCap Index (SPCBXM) indices in Panel C for robustness checks.[8] The average returns for AI-assets (non-AI-assets) in Panel A (B) range from 0.26% to 0.94% (0.10% to 0.13%) and exhibit a positively (negatively) skewed distribution. This initial finding indicates that despite a general downturn in the broader cryptocurrency market, AI-assets displayed exceptional resilience to the market trend.

## 3 Results

### 3.1 Baseline difference-in-difference estimates

We begin by estimating the traditional fixed-effects difference-in-difference (DID) model to provide preliminary estimates for the effects of the introduction of ChatGPT on AI-assets (outlined in Table 1, Panel A) as compared to corresponding non-AI asset control groups (outlined in Table 1, Panels B and C). The treatment groups GAI and CAI were exposed to the ChatGPT launch as of November 30, 2022. Models (1) to (6) in Table 2 reveal significant (mixed results) for ATTs of AI-assets in the two-month (one-month) period following the launch. The DID model necessitates the assumption of parallel trends in the pre-treatment phase for valid ATT identification. However, formal testing produces mixed results, with the rejection or non-rejection of the null hypothesis of parallel pre-treatment trends, contingent on model specification.[9]

---

[7] The SDID model necessitates a strictly balanced panel for estimation. Consequently, we excluded assets with missing or incomplete data from the analysis. Moreover, to ensure adequate liquidity, assets with daily trading volumes below $20,000 were excluded. SDID requires a greater number of treated assets than untreated assets, hence the selection of 15 (85) assets for GCKO (CMC), one less than the corresponding group GAI (CAI).

[8] The S&P Cryptocurrency Broad Digital Market (BDM) Index tracks the performance of crypto assets on exchanges that satisfy minimum liquidity and market capitalization criteria. It is designed to reflect the performance of the broad investable crypto universe. The S&P Cryptocurrency BDM Ex-LargeCap Index SPCBXL (Ex-MegaCap Index SPCBXM) excludes large (very large) market capitalization crypto assets from BDM accounting for around 78% (68%) of total market capitalization out of a potential 950+ assets considered by Lukka Prime. These metrics capture the performance of medium to small-size crypto assets. Data is obtained from spglobal.com.

---INSERT TABLE 2---

3.2 The synthetic difference-in-difference model

The inconclusive outcomes observed in the investigation of parallel trends necessitate the adoption of a more sophisticated analytical framework to ensure the veracity of model inferences and derived estimations. We continue by implementing the SDID methodology of Arkhangelsky et al. (2021) and Clarke et al. (2023) for determining the impact of a binary treatment variable (i.e., equal to one after the ChatGPT launch, $GPT_{it}$, and zero otherwise) on an outcome variable (i.e., crypto-asset returns, $R_{it}$) for a panel of $N$ crypto assets, observed over $T$ time intervals.[10] The procedure applies a treatment (i.e., ChatGPT launch) to be received

---

[9] Figure A1 provides some qualitative support for parallel pre-treatment trends using observed means and linear trend models.

[10] To summarize. We collect a strictly balanced panel of treated (untreated) AI-related (non-AI-related) crypto assets from CoinGecko denoted GAI (GCKO) and Coinmarketcap denoted CAI (CMC) with logged price returns, logged volumes and logged market capitalizations. A binary treatment variable is set equal to one for GAI and CAI after the ChatGPT launch. A two-way fixed-effects panel regression model is estimated where the outcome variable is crypto-asset log returns and the treatment variable is the binary treatment variable. The standard DID assigns equal weights to all time periods and groups, however SDID assigns optimally selected unit-specific weights. To succinctly summarize, pre-treatment residuals are calculated by regressing the outcome variable on a constant and time trend. The within-group covariance matrix of the pre-treatment residuals is calculated by taking the covariance of the pre-treatment residuals for each unit, and then averaging those covariances across all units. Weights are calculated by taking the inverse of the within-group covariance matrix, and then multiplying that by the pre-treatment residuals for each unit. The weights are then used to construct a weighted average of the

by units (i.e., AI-asset returns) in a block assignment. The objective of SDID is to consistently estimate causal effects of the treatment, even in the absence of a parallel trend assumption between the treatment and control units before the treatment event (i.e., ChatGPT launch). The average treatment effect (ATT) is estimated as a two-way fixed effects panel regression:

$$(\hat{\tau}^{sdid}, \hat{\mu}, \hat{\alpha}, \hat{\beta}) = \underset{\tau,\mu,\beta}{arg\ min} \left\{ \sum_{i=1}^{N} \sum_{t=1}^{T} (R_{it} - \mu - \alpha_i - \beta_i - GPT_{it}\tau)^2 \hat{\omega}_i^{sdid} \hat{\lambda}_t^{sdid} \right\} \quad (1)$$

where the estimand is generated from optimal unit $(\hat{\omega}_i^{sdid})$ and time $(\hat{\lambda}_t^{sdid})$ weights, the exclusion of which yields traditional DID estimation. Weight optimization ensures consistent comparison between treated units and control units through the application of similar trends before treatment. Weights are optimized with greater weighting to pre-treatment periods that are similar to post-treatment periods, to identify a constant difference between each control unit's post-treatment average and pre-treatment averages of all control units. The goal of weights is to balance pre- and post-treatment trends, with unit weights, $\hat{\omega}_i^{sdid}$, calculated by solving:

$$(\hat{\omega}_0, \hat{\omega}_i^{sdid}) = \underset{\hat{\omega}_0 \in \mathbb{R}, \omega \in \Omega}{arg\ min}\ \ell_{unit}(\omega_0, \omega)\ where \quad (2)$$

$$\ell_{unit}(\omega_0, \omega) = \sum_{t=1}^{T_{pre}} \left( \omega_0 + \sum_{i=1}^{N_{co}} \omega_0\, R_{it} - \frac{1}{N_{tr}} \sum_{i=N_{co}+1}^{N} R_{it} \right)^2 + \zeta^2 T_{pre} \|\omega\|_2^2,$$

$$\Omega = \left\{ \omega \in \mathbb{R}_+^N : \sum_{i=1}^{N_{co}} \omega_i = 1, \omega_i = N_{tr}^{-1}\ for\ all\ i = N_{co}+1, \ldots, N \right\},$$

---

outcomes of the untreated units, which is used as the estimate of the treatment effect. The weighted average is calculated by taking the sum of the products of the weights and the outcomes for each unit, and then dividing that sum by the sum of the weights. To further account for exogenous time-varying covariates, such as market capitalization and liquidity, the SDID model can be adjusted by regressing crypto-asset returns on these covariates.

where $\mathbb{R}_+$ denotes the positive real line and $\zeta$ denotes a regularization parameter matching the size of a typical outcome change of one period for unexposed units, multiplied by a scaling factor. The intercept term, $\omega_0$, allows greater flexibility in weight determination, relaxing requirements for unexposed and exposed pre-treatment trends to match perfectly, deeming parallel trends sufficient. This is due to use of fixed-effects, $\alpha_i$, which absorb constant differences in different units. The regularization penalty of Doudchenko and Imbens (2016) is further applied to enhance dispersion and ensure uniqueness of weights. Time weights, $\hat{\lambda}_t^{sdid}$, are not dependent on a regularization parameter, thus allowing correlated observations within time periods. They are implemented by solving:

$$(\hat{\lambda}_0, \hat{y}^{did}) = \underset{\lambda_0 \in \mathbb{R}, \lambda \in \Lambda}{\arg\min} \; \ell_{time}(\lambda_0, \lambda) \; where \tag{3}$$

$$\ell_{time}(\lambda_0, \lambda) = \sum_{i=1}^{N_{co}} \left( \lambda_0 + \sum_{t=1}^{T_{pre}} \lambda_0 R_{it} - \frac{1}{N_{tr}} \sum_{i=N_{co}+1}^{N} R_{it} \right)^2$$

$$\Lambda = \left\{ \lambda \in \mathbb{R}_+^T : \sum_{t=1}^{T_{pre}} \lambda_i = 1, \lambda_i = T_{post}^{-1} \; for \; all \; i = T_{pre} + 1, \ldots, T \right\}.$$

The SDID approach enables flexible estimation of both shared temporal aggregate factors and unit-specific factors, given the estimation of time fixed-effects ($\beta_i$) and unit fixed-effects ($\alpha_i$). To avoid multicollinearity, one $\alpha_i$ and one $\beta_i$ fixed-effect are normalized to zero. This has the advantage of seeking to match treated and control units on pre-treatment trends, rather than pre-treatment trends and levels, allowing for a constant difference between treatment and control units. Overall, SDID provides a flexible and robust approach for estimating ATT, by accounting for both shared temporal aggregate factors and unit-specific factors (Clarke et al., 2023). To further account for exogenous time-varying covariates of market capitalization ($Cap_{it}$) and liquidity ($Vol_{it}$) of crypto assets, the SDID model is adjusted as:

$$R_{it}^{res} = R_{it} - Cap_{it}\widehat{\beta_1} - Vol_{it}\widehat{\beta_2} \qquad (4)$$

where $\hat{\beta}$ is obtained from a regression of $R_{it}$ on $Cap_{it}$ and $Vol_{it}$ (Abadie et al., 2010).[11]

## 3.3 Synthetic difference-in-difference estimates

The explicit weighting mechanism applied in SDID ensures parallel trend identification prior to estimation. Table 3 reveals significant ATTs, ranging from 10.7% to 15.6%, for a narrow subset of AI-assets (GAI) surpassing counterpart non-AI-assets (GCKO) in the one-month window following the ChatGPT launch. ATTs increased in magnitude and significance, to between 35.5% and 41.3% in the two months following the launch. ATTs for a broader subset of AI-assets (CAI) were lower in magnitude, ranging from 5.7% to 6.9% (9.8% to 13.2%) where significant in the one-month (two-months) following the launch. Findings are robust to the models' specification, i.e., relative to benchmark cryptocurrency baskets (Panel A), relative to benchmark crypto indices (Panel B), and including covariates (Panel C). Results provide compelling evidence for the ChatGPT effect, resulting in a price increase of at least 10.7% (35.5%) beyond the general crypto market over temporal window of one-month (two-months) for a narrower group of AI-assets. Figure 1 illustrates the performance of models (1) and (2) from Table 3. Prior to the launch of ChatGPT, both AI-assets and non-AI-assets assets showed a high degree of synchronicity.[12] However, after the ChatGPT launch, we observed a significant increase in the divergence between these two groups, which continued to widen over the following two months.[13] Our results align with anecdotal reports by Jajric and Shen (2023) highlighting the surge in value of AI-related crypto assets (GAI) following the ChatGPT launch.

---

[11] A more detailed explanation of the SDID procedure is described in Arkhangelsky et al. (2021).

[12] The narrower group of AI-assets defined by coingecko.com (CAI) exhibit greater divergence in responses compared to the much broader group of AI-assets defined by coinmarketcap.com (GAI).

[13] The unit weights for these models are presented in Figure A2.

---INSERT TABLE 3---

---INSERT FIGURE 1---

3.4 Proxies for retail and institutional investor attention

Empirical evidence suggests that utilizing Google Trends search queries can serve as a pertinent proxy for gauging public attention directed towards specific topics, thereby potentially influencing investment decisions (e.g., Aslanidis et al., 2022; Dastgir et al., 2019; Philippas et al., 2019). To further explore this dynamic in the context of AI and ChatGPT attention, Table 4 (panels A and B) present panel fixed-effects regression estimates of Equation (5). $\alpha$ is the intercept. $e_t$ is the error term. $\Delta G_t$ in turn measures the change in daily search volume for the Google search terms (i) "AI", (ii) "Artificial Intelligence", and (iii) "ChatGPT" separately for each model. $D_t^{ChatGPT}$ is a dummy variable equal to one (zero) after (before) the ChatGPT launch. $D_t^{AI}$ is a dummy variable equal to one in panel (a) CoinGecko as AI-related or panel (b) Coinmarketcap as AI and Big Data-related, and zero otherwise.

$$r_t = \alpha + \begin{bmatrix} \left(\left((\beta_1(1-D_t^{AI})) + (\beta_2 D_t^{AI})\right)\left(1-D_t^{ChatGPT}\right)\right) \\ + \left(\left((\beta_3(1-D_t^{AI})) + (\beta_4 D_t^{AI})\right)D_t^{ChatGPT}\right) \end{bmatrix} \Delta G_t + e_t \quad (5)$$

---INSERT TABLE 4---

The response of non-AI-asset and AI-asset returns to changes in search volumes for respective search terms before the ChatGPT launch, denoted $\beta_1$ and $\beta_2$ respectively, is statistically insignificant for all six models. This indicates that proxies for retail attention to AI, such as Google search volume data of respective terms, were not perceived as significant pricing indicators for both non-AI ($\beta_1$) and AI ($\beta_2$) assets. However, after the ChatGPT launch, this dynamic changed markedly. When examining Panel A, which utilizes CoinGecko's categorization to more narrowly define AI-crypto assets, the response of non-AI ($\beta_3$) and AI ($\beta_4$) asset returns to the ChatGPT launch was statistically insignificant and significant respectively, across all three models. Changes in search volumes for these terms were deemed irrelevant for non-AI-assets ($\beta_3$) but highly relevant for investors in AI-assets ($\beta_4$). The positivity of the $\beta_4$ coefficient indicates that relative increases in searches for respective terms were associated with positive increases in AI-asset returns. Wald tests support statistically significant differences for the relevance of search terms as pricing indicators for AI-assets before and after the ChatGPT launch $[\beta_2 = \beta_4]$, as well as between non-AI and AI-assets after the ChatGPT launch $[\beta_3 = \beta_4]$. We yield lower magnitude responses utilizing Coinmarketcap's much broader categorization of AI and Big Data-related crypto assets. Our estimations lend empirical backing to observations by Redman (2023), which correlated record high Google Trend search queries for AI with the ChatGPT launch.

Institutional investors possess both the motivation and financial resources to respond rapidly to significant news (Ben-Rephael et al., 2017). To quantify institutional investor attention directed towards ChatGPT (AI), we develop a proxy that integrates both the frequency and sentiment of pertinent financial news wires found on Refinitiv's Thomson Reuters Eikon, a financial information platform extensively utilized by institutional investors.[14] By aggregating the daily count of news wires related to ChatGPT (AI) and adjusting it based on the sentiment polarity of the most widely read articles, we compute an index using a weighted average, which is subsequently normalized to a 0-100 range.[15] This index serves as a proxy for institutional investor attention, capturing both the magnitude and nuance of interest in ChatGPT (AI) on a daily basis.

---

[14] Refinitiv's Thomson Reuters Eikon (hereafter Eikon), a principal competitor to Bloomberg L.P., is a financial information platform that delivers real-time market data, news, analytics, and trading tools to financial professionals. The platform is predominantly employed by institutional investors, including banks, asset managers, hedge funds, and other financial institutions, as a basis for their investment decisions. Financial news wires are typically accessed by institutional investors, financial professionals, and corporations and businesses. As of 2023, Eikon accommodates over 400,000 end users across 190 countries (Eikon, 2023).

The reaction of non-AI-asset and AI-asset returns to changes in the proxy for institutional investor attention to ChatGPT (AI) financial news wires in Panel C (Panel D) before the ChatGPT launch, denoted $\beta_1$ and $\beta_2$ respectively, and non-AI-assets after the ChatGPT launch, denoted $\beta_3$ is statistically insignificant across all four models.[16] This suggests that the sentiment and volume of financial news wires related to these terms were not considered significant pricing indicators for institutional investors. Intriguingly, the impact on AI-assets following the ChatGPT launch denoted $\beta_4$ exhibits statistical significance in certain instances; however, the magnitude of response is considerably smaller than that observed in Panel A. This finding implies that the demand for AI-assets after the ChatGPT launch was predominantly driven by retail investors. Our findings concur with a JP Morgan survey, ascertaining that a sizeable 72% of institutional investors harbored no intentions to engage in cryptocurrency trading, with a mere 14% intending to trade crypto assets in the next five years (Ozawa, 2023).

## 4 Conclusion

This paper demonstrates the ChatGPT launch had a significant impact on the performance of AI-related crypto-assets, despite the overall cryptocurrency market being in a bearish state and risk-averse investor appetites. Using synthetic difference-in-differences, we found average price increases of at least 10.7% (35.5%) in the one-month (two-month) period following the launch. Firstly, we contribute to the literature on the impact of technology launches on financial markets and the role of sentiment and hype in shaping market outcomes. Secondly, we demonstrate the launch of an innovative technology, ChatGPT, can benefit from widespread publicity, which, based on proxies of investor attention from Google search volume data, has spillover effects on investor's perceptions of ChatGPT and AI potential, resulting in higher AI-asset valuations. In the context of signaling theory, this supports the idea that (social) media coverage can proxy quality signals for investors by underscoring the (perceived) potential of AI technologies and AI-sector opportunities, illuminated in this paper via the effect of the ChatGPT launch on AI-assets. Further studies should consider spillover effects from pricing of AI-related stocks to AI-related crypto assets; the role of influencers such as Elon Musk in promoting ChatGPT; and the informational transmission mechanisms by which AI-mania propagated.

---

[15] The sentiment measure for each article is demarcated by Eikon along a continuum, extending from Mostly Negative (-1) through Balanced (0) to Mostly Positive (+1).

[16] Eikon categorises financial news wires pertaining to Artificial Intelligence and AI within the same classification, hence the estimates cannot be separated.

**Tables and Figures**

**Table 1. Descriptive statistics for crypto asset returns**

| Ticker | Name | Obs | N | Mean | SD | Min | Max | Skew | JB |
|---|---|---|---|---|---|---|---|---|---|
| *Panel A: AI-related crypto asset basket categories* | | | | | | | | | |
| GAI | CoinGecko: Artificial Intelligence | 2,000 | 16 | 0.0094 | 0.092 | -0.44 | 0.59 | 1.10 | 0.00*** |
| CAI | Coinmarketcap: AI and Big Data | 10,750 | 86 | 0.0026 | 0.092 | 1.54 | 1.53 | 0.78 | 0.00*** |
| *Panel B: Non-AI, non-stablecoin comparison control groups* | | | | | | | | | |
| GCKO | CoinGecko: Control Group | 1,865 | 15 | 0.0013 | 0.045 | -0.55 | 0.37 | -0.53 | 0.00*** |
| CMC | Coinmarketcap: Control Group | 10,625 | 85 | 0.0010 | 0.048 | -0.55 | 0.37 | -0.22 | 0.00*** |
| *Panel C: Non-stablecoin comparison index control groups* | | | | | | | | | |
| SPCBXL | S&P Crypto BDM Ex-LargeCap Index | 90 | 500+ | 0.0006 | 0.037 | -0.15 | 0.10 | -0.97 | 0.00*** |
| SPCBXM | S&P Crypto BDM Ex-MegaCap Index | 90 | 500+ | 0.0004 | 0.032 | -0.16 | 0.10 | -1.21 | 0.00*** |

*Notes: The table reports group statistics for continuously compounded returns for defined asset groupings, over the October 01, 2022 to January 31, 2023 period, comprising 125 periods for Panels A and B, and 90 periods for Panel C attributable to the fact that trading only occurs on working days. JB denotes the Jarque-Bera test for normality. *, **, *** indicate significance at the 10%, 5% and 1% levels. GAI includes FET, AGIX, ALI, NMR, FITFI, SDAO, VXV, ORAI, GNY, DBC, MOOV, HERA, BOTTO, MAN, RAVEN, EFX. GCKO includes BTC, ETH, BNB, XRP, ADA, DOGE, MATIC, SOL, DOT, SHIB, LTC, AVAX, TRON, UNI, ATOM. CAI includes GRT, AGIX, ROSE, FET, OCEAN, RLC, ALI, NMR, DKA, PHA, CQT, LAT, CTXC, SDAO, PHB, VXV, MDT, DIA, GNY, DATA, UPP, PRQ, XMON, DX, PRE, DBC, AION, DOCK, BDP, GOC, KAT, ORAI, HAI, SWASH, LBC, MAN, BOTTO, ZCN, SIDUS, XRT, MOOV, RAVEN, GLQ, AGRS, FOAM, CIRUS, ROOBEE, TRV, EFX, LAMB, AXIS, SAN, LIME, SENATE, EPK, SEELE, CND, IDNA, BCUBE, ANW, AIRI, CPC, ALBT, MARSH, ARCONA, TRAVA, RAZE, DDD, UTU, OCE, UPI, BTO, DTA, JAR, LML, ZEBI, CNTM, UBEX, ZSC, CYL, ASTO, XETA, OJA, DHX, SNS. CMC includes BTC, ETH, BNB, XRP, ADA, DOGE, MATIC, SOL, DOT, SHIB, LTC, AVAX, TRX, UNI, ATOM, LINK, LEO, ETC, XMR, TON, OKB, BCH, XLM, LDO, NEAR, APE, CRO, FIL, ALGO, HBAR, VET, QNT, ICP, FTM, MANA, BIT, SAND, AXS, AAVE, EOS, THETA, FLOW, EGLD, XTZ, LUNC, CHZ, FXS, HT, KCS, BSV, IMX, CAKE, CRV, MKR, ZEC, XEC, BTT, DASH, SNX, MIOTA, MINA, NEO, OP, TWT, KLAY, RUNE, OSMO, LRC, PAXG, ENJ, RPL, ZIL, GT, DYDX, CVX, CSPR, 1INCH, BAT, LUNA, RNDR, NEXO, STX, HNT, COMP.*

**Table 2. DID estimation results for returns**

| Model | AI Category | Controls | Covariates | ATT (0 to 1 Month) | ATT (0 to 2 Months) | Parallel Trends |
|---|---|---|---|---|---|---|
| *Panel A: Baseline Models* | | | | | | |
| (1) | GAI | GCKO | - | 0.08481 (0.08730) | 0.32139*** (0.10713) | [0.27940] |
| (2) | CAI | CMC | - | 0.01563 (0.03778) | 0.07513 (0.04569) | [0.02730] |
| *Panel B: Robustness Checks with Index Controls* | | | | | | |
| (3) | GAI | SPCBXL | - | 0.15600* (0.07367) | 0.41309*** (0.10010) | [0.78020] |
| (4) | GAI | SPCBXM | - | 0.14022* (0.07367) | 0.38885*** (0.10010) | [0.29490] |
| (5) | CAI | SPCBXL | - | 0.03974 (0.03211) | 0.12219*** (0.04107) | [0.10550] |
| (6) | CAI | SPCBXM | - | 0.02395 (0.03211) | 0.09796** (0.04107) | [0.00020] |
| *Panel C: Robustness Checks with Covariates* | | | | | | |
| (7) | GAI | GCKO | ln(vol) & ln(cap) | -0.04811 (0.03514) | -0.05268 (0.04845) | [0.08790] |
| (8) | GAI | GCKO | ln(vol) | 0.02531 (0.08130) | 0.17657 (0.10673) | [0.46970] |
| (9) | GAI | GCKO | ln(cap) | -0.04113 (0.03097) | -0.04608 (0.04464) | [0.09990] |
| (10) | CAI | CMC | ln(vol) & ln(cap) | 0.00494 (0.01533) | 0.02712 (0.02120) | [0.91170] |
| (11) | CAI | CMC | ln(vol) | 0.01200 (0.03604) | 0.06232 (0.03977) | [0.03700] |
| (12) | CAI | CMC | ln(cap) | 0.00740 (0.01600) | 0.03089 (0.02228) | [0.89180] |

*Notes: The table reports average treatment effects (ATTs) for a panel fixed-effects model using a group of crypto assets defined as 'AI'-related by CoinGecko (GAI) in Panel A and 'AI and group of Big Data'-related crypto assets by Coinmarketcap (CAI) in Panel B over the October 01, 2022 to January 31, 2023 period. The treatment groups GAI and CAI were exposed to the ChatGPT launch, as of November 30, 2022, implying $T_{pre} = 60$ pre-treatment periods and $T_{post} = 66$ post-treatment periods. Covariates consists of log-transformed trading volumes, ln(vol), and log-transformed market capitalizations, ln(cap). P-values from Parallel Trends tests (F-statistics) appear in square brackets. Bias-corrected cluster-robust standard errors are reported in parentheses (Bell and McCaffrey, 2003). \*, \*\*, \*\*\* indicate significance at the 10%, 5% and 1% levels.*

**Table 3. SDID estimation results for returns**

| Model | AI Category | Controls | Covariates | ATT (0 to 1 Month) | ATT (0 to 2 Months) |
|---|---|---|---|---|---|
| *Panel A: Baseline Models* | | | | | |
| (1) | GAI | GCKO | - | 0.11151* (0.06653) | 0.36416*** (0.09432) |
| (2) | CAI | CMC | - | 0.05714** (0.02389) | 0.11128*** (0.03490) |
| *Panel B: Robustness Checks with Index Controls* | | | | | |
| (3) | GAI | SPCBXL | - | 0.15600** (0.07380) | 0.41309*** (0.10171) |
| (4) | GAI | SPCBXM | - | 0.14022*** (0.07380) | 0.38885*** (0.10171) |
| (5) | CAI | SPCBXL | - | 0.03974 (0.03224) | 0.12219*** (0.04153) |
| (6) | CAI | SPCBXM | - | 0.02395 (0.03224) | 0.09796** (0.04153) |
| *Panel C: Robustness Checks with Covariates* | | | | | |
| (7) | GAI | GCKO | ln(vol) & ln(cap) | 0.10709*** (0.06576) | 0.35494*** (0.09264) |
| (8) | GAI | GCKO | ln(vol) | 0.10731* (0.09293) | 0.35579*** (0.09293) |
| (9) | GAI | GCKO | ln(cap) | 0.11129* (0.06639) | 0.36325*** (0.09401) |
| (10) | CAI | CMC | ln(vol) & ln(cap) | 0.06633*** (0.02350) | 0.12667** (0.03374) |
| (11) | CAI | CMC | ln(vol) | 0.06672*** (0.02362) | 0.12827*** (0.03411) |
| (12) | CAI | CMC | ln(cap) | 0.06855*** (0.02368) | 0.13254*** (0.03441) |

*Notes: The table reports synthetic difference-in-difference estimates based on Arkhangelsky et al. (2021), i.e., average treatment effects (ATTs), using a group of crypto assets defined as 'AI'-related by CoinGecko (GAI) in Panel A and 'AI and Big Data'-related crypto assets by Coinmarketcap (CAI) in Panel B over the October 01, 2022 to January 31, 2023 period. The treatment groups GAI and CAI were exposed to the ChatGPT launch, as of November 30, 2022, implying $T_{pre} = 60$ pre-treatment periods and $T_{post} = 66$ post-treatment periods. Covariates consists of log-transformed trading volumes, ln(vol), and log-transformed market capitalizations, ln(cap). Standard errors in parentheses are based on 500 bootstrapped replications. *, **, *** indicate significance at the 10%, 5% and 1% levels.*

**Table 4: Response of crypto assets to search volume, controlling for the ChatGPT launch and AI-centric assets**

| Search Term | Obs. | α | $\beta_1$ | $\beta_2$ | $\beta_3$ | $\beta_4$ | [$\beta_1=\beta_2$] | [$\beta_3=\beta_4$] | [$\beta_1=\beta_3$] | [$\beta_2=\beta_4$] | Adj. $R^2$ |
|---|---|---|---|---|---|---|---|---|---|---|---|
| *Panel A: CoinGecko (Retail Investors)* | | | | | | | | | | | |
| "AI" | 4,305 | 0.03*** (0.01) | -0.04 (0.05) | -0.04 (0.06) | 0.00 (0.03) | 0.09*** (0.03) | [0.97] | [0.02]** | [0.47] | [0.03]** | 0.02 |
| "Artificial Intelligence" | 4,305 | 0.03*** (0.01) | -0.03 (0.04) | -0.03 (0.05) | 0.00 (0.02) | 0.07*** (0.02) | [0.93] | [0.01]** | [0.58] | [0.05]** | 0.02 |
| "ChatGPT" | 4,305 | 0.03*** (0.01) | -0.06 (0.18) | -0.01 (0.20) | -0.02 (0.02) | 0.11*** (0.02) | [0.86] | [0.00]*** | [0.87] | [0.53] | 0.09 |
| *Panel B: Coinmarketcap (Retail Investors)* | | | | | | | | | | | |
| "AI" | 21,279 | -0.12*** (0.00) | 0.00 (0.02) | -0.02 (0.02) | 0.03** (0.01) | 0.02* (0.01) | [0.44] | [0.80] | [0.35] | [0.09]* | 0.03 |
| "Artificial Intelligence" | 21,279 | -0.12*** (0.00) | 0.00 (0.02) | -0.02 (0.02) | 0.02 (0.01) | 0.01 (0.01) | [0.47] | [0.93] | [0.51] | [0.13] | 0.02 |
| "ChatGPT" | 21,279 | -0.12*** (0.00) | -0.01 (0.08) | -0.01 (0.08) | 0.00 (0.01) | 0.01* (0.01) | [0.98] | [0.23] | [0.88] | [0.79] | 0.02 |
| *Panel C: CoinGecko (Institutional Investors)* | | | | | | | | | | | |
| "ChatGPT" | 4,305 | 0.03*** (0.01) | -0.06 (0.18) | -0.01 (0.20) | 0.00 (0.00) | 0.01* (0.01) | [0.86] | [0.16] | [0.76] | [0.92] | 0.00 |
| "Artificial Intelligence" | 4,305 | 0.03*** (0.01) | 0.00 (0.00) | 0.00 (0.00) | 0.00 (0.00) | 0.01** (0.01) | [0.98] | [0.21] | [0.43] | [0.04]** | 0.00 |
| *Panel D: Coinmarketcap (Institutional Investors)* | | | | | | | | | | | |
| "ChatGPT" | 21,279 | -0.12*** (0.00) | -0.01 (0.08) | -0.01 (0.08) | 0.00 (0.00) | 0.00 (0.00) | [0.98] | [0.50] | [0.88] | [0.89] | 0.00 |
| "Artificial Intelligence" | 21,279 | -0.12*** (0.00) | 0.00 (0.00) | 0.00 (0.00) | 0.00 (0.00) | 0.01** (0.00) | [0.82] | [0.52] | [0.19] | [0.03]** | 0.00 |

*Notes: The table reports panel fixed-effects regression estimates of Equation (5) (i.e., $r_t = \alpha + [\beta_1(1-D_t^{ChatGPT})(1-D_t^{AI}) + \beta_2(1-D_t^{ChatGPT})(D_t^{AI}) + \beta_3(D_t^{ChatGPT})(1-D_t^{AI}) + \beta_4(D_t^{ChatGPT})(D_t^{AI})]\Delta G_t + e_t)$ with White robust cross-sectional standard errors. $r_t$ denotes crypto asset returns. $\Delta G_t$ denotes the change in Google search volume for the respective term. $D_t^{ChatGPT}$ is a dummy variable equal to one (zero) after (before) the ChatGPT launch on November 30, 2022. $D_t^{AI}$ is a dummy variable indicating whether a crypto asset is defined as "AI"-related by CoinGecko in Panel A and "AI and Big Data"-related crypto assets by Coinmarketcap in Panel B over the October 01, 2022 to January 31, 2023 period. Standard errors appear in parentheses. P-values from Wald tests (F-statistics) appear in square brackets. \*, \*\*, \*\*\* indicate significance at the 10%, 5% and 1% levels.*

**Figure 1. Outcome trends**

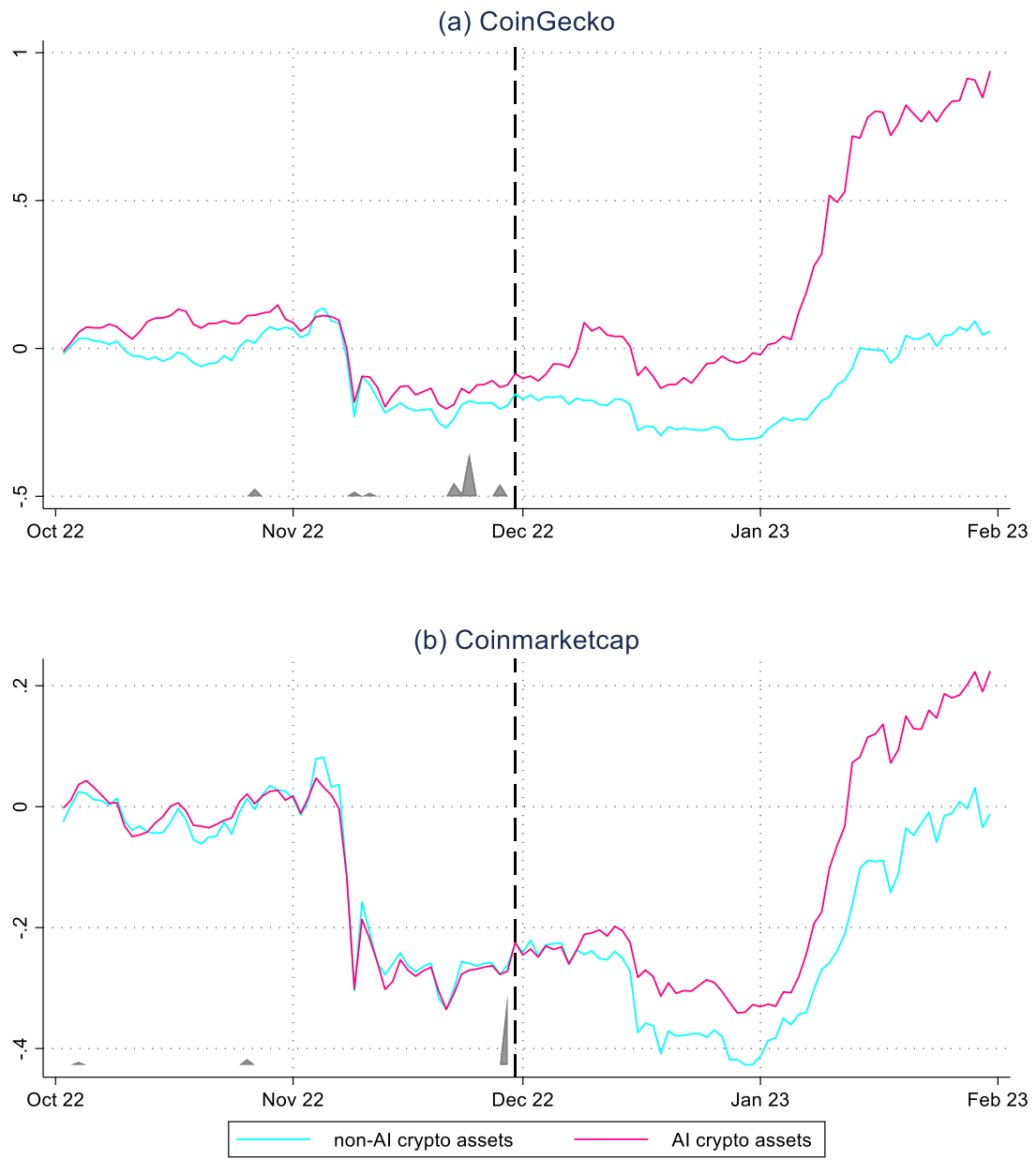

*Notes: Fig.1 panels (a) and (b) show the outcome trends of models (1) and (2) respectively from Table 3 following the methodology of Arkhangelsky et al. (2021). The dashed line indicates the ChatGPT launch on November 30, 2022.*

# Appendix

**Figure A.1. Graphical diagnostics for parallel trends**

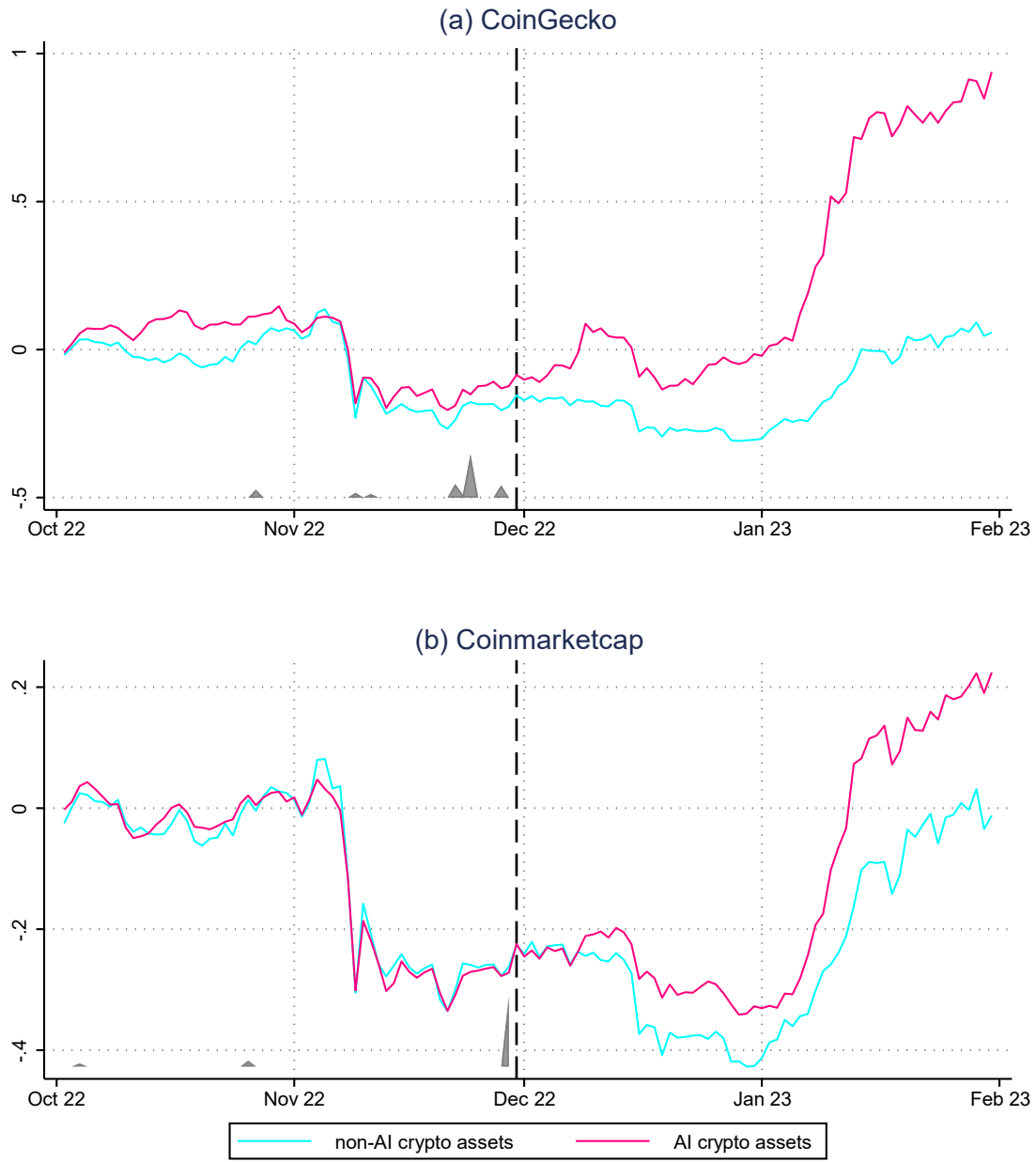

*Notes: Fig.A.1 shows diagnostic graphs for inference of parallel trends using model (1) from Table 2, with very similar results obtained for models (2) to (5). The observed means (a) shows the average outcome over time for the treatment and control groups. The linear-trends model (b) augments the DID model to include interactions for time and treatment, plotting predicted values for treatment and control groups. The dashed line indicates the ChatGPT launch on November 30, 2022.*

**Figure A.2. Unit weights for SDID models on returns**

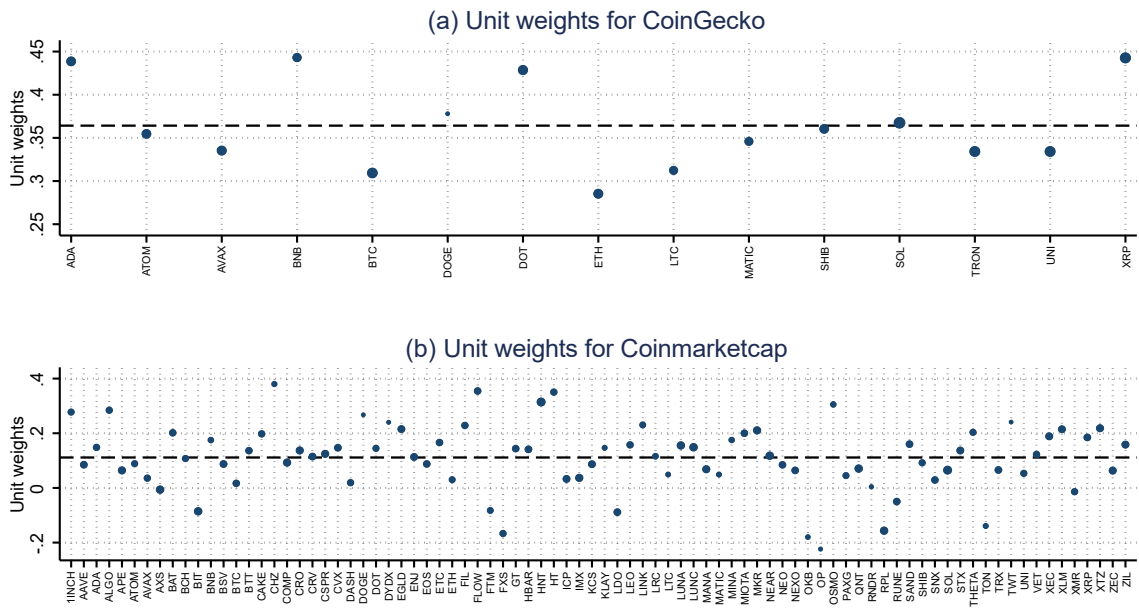

*Notes: Fig.A2 panels (a) and (b) show the unit weights of models (1) and (2) respectively from Table 3 following the methodology of Arkhangelsky et al. (2021).*